\documentclass{article}
\usepackage{spconf}
\usepackage{amsmath}
\usepackage{graphicx}
\usepackage{cite}
\usepackage{mathptmx}  
\usepackage{multirow}
\usepackage{array}
\usepackage{booktabs}
\usepackage{subcaption}
\usepackage{colortbl}
\usepackage{amssymb}
\usepackage[utf8]{inputenc}
\usepackage[normalem]{ulem}
\usepackage{comment}
\usepackage{balance}
\usepackage{siunitx}
\usepackage{hyperref}
\usepackage{url}
\usepackage{pifont}

\usepackage{soul}

\newcommand{\cmark}{\ding{51}}
\newcommand{\xmark}{\ding{55}}

\title{Reliable COVID-19 Detection using Chest X-Ray Images}

\name{Aysen Degerli$^{\dagger}$, Mete Ahishali$^{\dagger}$, Serkan Kiranyaz$^{\ast}$, Muhammad E. H. Chowdhury$^{\ast}$, and Moncef Gabbouj $^{\dagger}$}
  
  \address{$^{\dagger}$ Faculty of Information Technology and Communication Sciences, Tampere University, Tampere, Finland\\
      $^{\ast}$ Department of Electrical Engineering, Qatar University, Doha, Qatar}

\begin{document}
\ninept
\maketitle

\begin{abstract}
Coronavirus disease 2019 (COVID-19) has emerged the need for computer-aided diagnosis with automatic, accurate, and fast algorithms. Recent studies have applied Machine Learning algorithms for COVID-19 diagnosis over chest X-ray (CXR) images. However, the data scarcity in these studies prevents a reliable evaluation with the potential of overfitting and limits the performance of deep networks. Moreover, these networks can discriminate COVID-19 pneumonia usually from healthy subjects only or occasionally, from limited pneumonia types. Thus, there is a need for a robust and accurate COVID-19 detector evaluated over a large CXR dataset. To address this need, in this study, we propose a reliable COVID-19 detection network: ReCovNet, which can discriminate COVID-19 pneumonia from 14 different thoracic diseases and healthy subjects. To accomplish this, we have compiled the largest COVID-19 CXR dataset: QaTa-COV19 with 124,616 images including 4603 COVID-19 samples. The proposed ReCovNet achieved a detection performance with 98.57\% sensitivity and 99.77\% specificity.
\end{abstract}


\section{Introduction}
Coronavirus disease 2019 (COVID-19), caused by severe acute respiratory syndrome Coronavirus-2 (SARs-CoV-2), was declared a pandemic by the World Health Organization in March 2020. The disease affects seriously people in high-risk groups (especially the elderly) leading to hospitalization, intubation, and even death \cite{world2020coronavirus}. In order to prevent the spread of the disease, detection, and isolation of infected patients have the utmost importance. However, the diagnosis of COVID-19 is challenging due to its similar symptoms with other viral infections such as fever, cough, fatigue, and breathlessness \cite{singhal2020review}. Therefore, reliable detection of the disease has significant importance.

Recent diagnostic tools to detect COVID-19 are nucleic acid detection with real-time polymerase chain reaction (RT-PCR), computed tomography (CT), and chest X-ray (CXR) imaging. RT-PCR has become the gold standard for COVID-19 diagnosis. However, RT-PCR tests suffer from instability and high false alarm rate \cite{tahamtan2020real}. On the other hand, CT imaging has higher sensitivity compared to RT-PCR test; thus, recommended for the suspected cases \cite{bernheim2020chest}. However, the performance of CT imaging in the early COVID-19 cases has limited sensitivity \cite{bernheim2020chest}. Thus, CXR imaging is widely used for the diagnosis of COVID-19 mainly because of its advantages that are faster acquisition, less radiation exposure, and easy accessibility compared to the aforementioned tools \cite{brenner2007computed}. 

Many studies utilized Deep Learning (DL) algorithms for COVID-19 detection \cite{ chowdhury2020pdcovidnet, pham2020classification, chowdhury2020can}. However, the reliability of these models is under question due to their hidden decision-making process. In fact, the activation maps of the deep models reveal the unreliability of their decision-making process, where irrelevant areas on the CXRs, outside of the lung area such as bones, background, or text, affect the decision of the network. Therefore, several studies \cite{alom2020covidmtnet, haghanifar2020covid, goldstein2020covid} attempted to prevent deep models to learn from these irrelevant areas on the CXRs with a two-staged approach for COVID-19 detection by processing only the lung areas with lung segmentation as their first stage. At the second stage, only the segmented lung area on the CXRs are given to the deep models as the input. Although these studies have achieved good performance for COVID-19 detection, data scarcity is the main drawback that can yield overfitting and hinders an accurate evaluation. Moreover, the datasets used in these studies encapsulate none or limited thoracic diseases, i.e., viral and bacterial pneumonia against COVID-19 pneumonia that makes them unreliable in real-case scenarios for COVID-19 diagnosis. 

In this study, to address the aforementioned issues we propose ReCovNet: a \textbf{re}liable \textbf{COV}ID-19 detection \textbf{net}work, which is an end-to-end network solution. Instead of detecting COVID-19 directly from the CXR image or the segmented lung area on the CXR, we embed this information into the ReCovNet model by transfer learning from a segmentation network. For this purpose, we initially train the segmentation network and detach its encoder block to reconstruct the ReCovNet model for COVID-19 detection. Additionally, in this work, we extend the QaTa-COV19 dataset that was introduced in our previous study \cite{degerli2020covid19}. The extended version of QaTa-COV19 is the largest COVID-19 dataset with $124,616$ images including $4603$ COVID-19 samples. The control group CXRs consists of $14$ different thoracic diseases and healthy subjects. Moreover, the QaTa-COV19 consists of a subset of $1065$ early COVID-19 cases showing no or limited sign of COVID-19 pneumonia, which makes the diagnosis more challenging. Accordingly, the proposed ReCovNet trained over the largest QaTa-COV19 dataset has an outstanding performance with a reliable diagnosis compared to state-of-the-art deep models. Lastly, the benchmark QaTa-COV19 dataset is publicly shared with the research community\footnote{The benchmark QaTa-COV19 is publicly shared at the repository \href{https://www.kaggle.com/aysendegerli/qatacov19-dataset}{https://www.kaggle.com/aysendegerli/qatacov19-dataset}.}.
 
The rest of the paper is organized as follows. In Section \ref{sec:methods}, we introduce the QaTa-COV19 dataset and give the details of our proposed ReCovNet model along with the state-of-the-art deep models. In Section \ref{sec:experiments}, we report the experimental results, and we conclude the paper in Section \ref{sec:conclusion}.

\section{Materials and Methodology}\label{sec:methods}
In this section, first we introduce the benchmark QaTa-COV19 dataset. Then, the state-of-the-art deep models are introduced for COVID-19 diagnosis. Lastly, we propose the ReCovNet model for reliable COVID-19 detection.

\begin{figure*}[t!]
    \centering
    \includegraphics[width=0.85\linewidth]{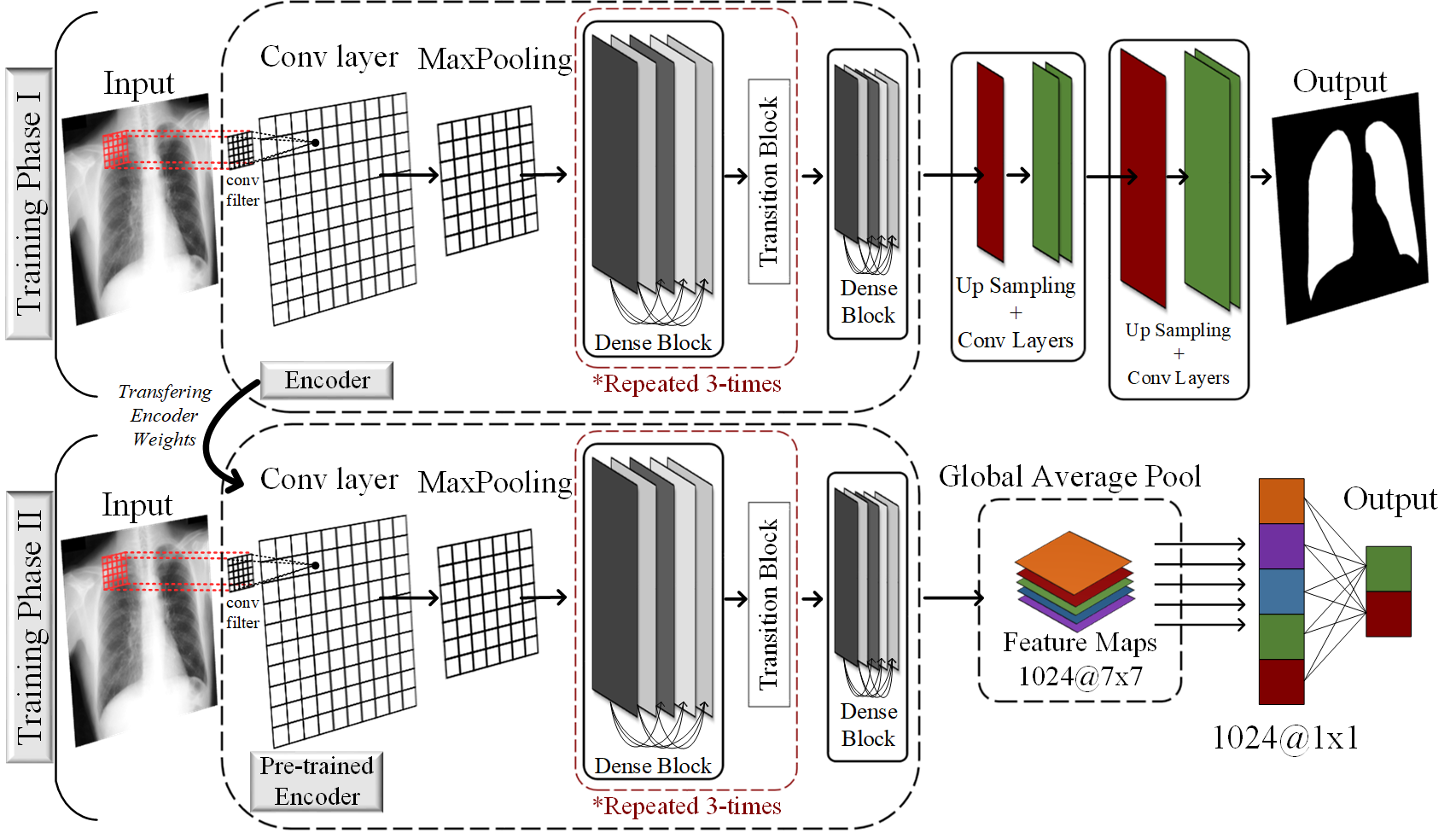}
    \caption{The proposed ReCovNet, where the transfer learning is performed from the segmentation network initially trained on CXRs for lung segmentation.}
    \label{fig:segmentation}
\end{figure*}

\subsection{The Benchmark QaTa-COV19 Dataset}
The benchmark QaTa-COV19 dataset, compiled by researchers of Qatar University and Tampere University is so far the largest COVID-19 dataset including $4603$ COVID-19 and $120,013$ control group CXRs. The detection task on this dataset is especially challenging since QaTa-COV19 consists of $1065$ samples from early COVID-19 cases that show no or limited sign of COVID-19 pneumonia. COVID-19 samples have been collected from publicly available datasets and repositories \cite{vaya2020bimcv, covidimage, coviddatabase, cohen2020covid, radiodatabase, chestimaging, haghanifar2020covid}, and were preprocessed by excluding low-quality images and any duplication. The control group images were collected from several datasets: ChestX-ray14 \cite{wang2017chestx}, X-rays from pediatric patients \cite{kermany2018identifying}, and Chest X-rays (Indiana University) \cite{indianachest}. We have only used the bacterial and viral pneumonia CXRs from pediatric patients to increase pneumonia samples for a challenging diagnosis. Additionally, we included only the lateral-view CXRs from Chest X-rays (Indiana University) dataset since all other samples in the control group are from frontal-view CXRs, whereas COVID-19 samples include CXRs both from lateral and frontal views. 

\begin{table}[ht!]
\centering
\caption{Details of QaTa-COV19 dataset.}
\resizebox{.48\textwidth}{!}{
\begin{tabular}{|c|c|c|c|c|}
\hline
\rowcolor[gray]{.90}Data & \begin{tabular}[c]{@{}c@{}}Training \\ Samples\end{tabular} & \begin{tabular}[c]{@{}c@{}}Augmented\end{tabular} & \begin{tabular}[c]{@{}c@{}}Augmented \\ Training Samples\end{tabular} & \begin{tabular}[c]{@{}c@{}}Test \\ Samples\end{tabular} \\ \hline \hline

\begin{tabular}[c]{@{}c@{}}ChestX-ray14\end{tabular} & $86,524$ & \xmark & $86,524$ & $25,596$ \\ \hline

\rowcolor[gray]{.95}\begin{tabular}[c]{@{}c@{}}Bacterial \\ Pneumonia\end{tabular} & $2130$ & \cmark & $5000$ &  $630$\\  \hline

\begin{tabular}[c]{@{}c@{}}Chest X-rays \\ (Indiana University)\end{tabular} & $2816$ & \cmark & $5000$ & $832$ \\ \hline 

\rowcolor[gray]{.95}\begin{tabular}[c]{@{}c@{}}Viral \\ Pneumonia\end{tabular} & $1146$ & \cmark & $5000$ & $339$ \\ \hline \hline

\begin{tabular}[c]{@{}c@{}}COVID-19 \\ \end{tabular} & $3553$ & \cmark & $10,000$ & $1050$ \\  \hline \hline

Total & 96,169 &  & $\textbf{111,524}$ & $\textbf{28,447}$ \\ \hline
\end{tabular}}
\label{tab:numberofsamples}
\end{table}

Table \ref{tab:numberofsamples} shows the number of samples in the QaTa-COV19 dataset. COVID-19 detection is performed against the control group images, which consists of $14$ different thoracic diseases and healthy subjects. Therefore, we perform a binary classification problem. Since the train and test sets of the ChestX-ray14 dataset are predefined, we have randomly split Chest X-rays (Indiana University), bacterial and viral pneumonia, and COVID-19 CXRs with the same train/ test ratio as in \cite{wang2017chestx}. The CXRs in the dataset are resized to $224\times224$ pixels. We have augmented the images except for ChestX-ray14 samples using the Image Data Generator in Keras. The images are $10\%$ randomly shifted both horizontally and vertically, and in a $10$-degree range randomly rotated. Lastly, the '\textit{nearest}' mode is selected to fill the blank sections. 

\subsection{COVID-19 Detection with Deep Models}
DL algorithms achieved state-of-the-art results on many computer vision tasks, including COVID-19 detection. Especially during the pandemic, recent studies concluded that DL algorithms with Convolutional Neural Networks can achieve outstanding performance for COVID-19 diagnosis. Nevertheless, the major issue in DL is that supervised deep models require a large amount of data to generalize well over unseen data. Thus, when subjected to data scarcity, such models fail in the testing phase due to overfitting. In this study, our first objective is to investigate the performances of state-of-the-art deep models by transfer learning on the largest COVID-19 dataset: QaTa-COV19. The state-of-the-art networks are selected as follows:
\begin{itemize}
    \item \textbf{DenseNet-121} \cite{huang2017densely} is a $121$-layer deep network that achieves a maximum information flow by connecting the layers with additional input nodes.
    \item \textbf{ResNet-50} \cite{he2016deep} is a deep network with $50$-layers that introduces residual blocks to prevent gradient vanishing in deep model structures by shortcut connections that merge input and output through the stacked layers.
    \item \textbf{Inception-v3} \cite{szegedy2016rethinking} is a deep network with low computational complexity compared to other state-of-the-art deep models. The reduced complexity is ensured by pruning and factorizing operations inside the network.
    \item \textbf{Inception-ResNet-v2} \cite{szegedy2017inception} unites the structure of the inception model \cite{szegedy2016rethinking} with residual blocks \cite{he2016deep} to achieve state-of-the-art results in computer vision tasks with a less computational cost.
\end{itemize}
In order to utilize the deep models in the COVID-19 detection task, we modify their output layers by inserting a global average pooling layer, a fully connected layer with 2-neurons, and a softmax activation function. The transfer learning is performed on the models by initializing their weights with the ImageNet weights. 

\subsection{ReCovNet: Reliable COVID-19 Detection Network}
DL algorithms are often considered as black-box since their decision-making process is latent. In order to reveal their mystery in the decision-making process, the authors in \cite{selvaraju2017grad} proposed Grad-CAM method that computes activation maps indicating the areas on the input image considered by the deep model during the classification task. In the COVID-19 detection task, our observations on the activation maps with the Grad-CAM approach show that the state-of-the-art deep models tend to learn and perform the classification from irrelevant areas on the CXRs, such as bones, background, or text. Therefore, the decisions of these models may be considered unreliable for COVID-19 detection. In order to overcome the unreliability issue, this study proposes ReCovNet: an end-to-end network for reliable COVID-19 detection.

ReCovNet is a deep network that considers the lung areas on the input CXR images to detect COVID-19 pneumonia. The structure of the proposed ReCovNet is given in Fig. \ref{fig:segmentation}. Accordingly, to construct ReCovNet, a segmentation network is trained in phase-I. The structure of the lung segmentation network is a convolutional autoencoder that maps the input image, $\mathbf{X}$ to its corresponding output mask, $\mathbf{M}$: $\mathbf{M} \xleftarrow[]{} P_{\theta,\phi}(\mathbf{X})$. Any deep model can be used as the encoder block of the network, $\varepsilon_{\theta}$. On the other hand, the decoder block of the segmentation network is similar to the U-Net \cite{ronneberger2015u} model except for its u-shaped architecture, where the low-level features at the encoder block are concatenated with the high-level features at the decoder level. The u-shaped architecture is excluded by removing the skip connections, which performs the concatenation operation. The reason for constructing an encoder-decoder network without skip connections is that the contributions from the initial layers are avoided; therefore, the network can make decisions from the high-level features that are closer to segmentation mapping of the input image. Based on our observations, this approach improves the performance of ReCovNet in terms of reliability observed in the activation maps. The decoder block of the segmentation network consists of $\phi \in \{b_j, w_j\}_{j=1}^L$ with $L$ number of layers composed of five stages. Each stage consists of an upsampling layer by $\times2$, and sequentially two times of the convolutional layer, batch normalization, and Rectified Linear Unit (ReLU) activation function. The output of the last stage is connected to a convolutional layer with a sigmoid activation function to reconstruct the segmentation mask at the output. In order, the number of convolutional layer filters are $\{256, 128, 64, 32, 16, 1\}$ with kernel of size of $k=(3\times3)$. Lastly, training is performed over $N$ number of samples $\{x_{s,train}^j, M^j\}_{j=1}^{N}$, where $x_s$ and $M$ are the training data, and ground-truth segmentation masks, respectively. The loss function used in training is a hybrid function, which is the summation of the binary focal and dice loss functions. 

During phase-II of the training, we construct the convolutional layers of ReCovNet by $\varepsilon_{\theta}$ that generates the latent features $f \xleftarrow[]{} \varepsilon_{\theta}$(.). Then, $f$ is vectorized and downsampled by attaching a global average pooling layer and a fully connected layer with $2$-neurons using softmax activation function. We perform the classification task with categorical cross-entropy loss function by training ReCovNet over $N$ number of samples $\{x_{train}^j, y_{train}^j\}_{j=1}^{N}$, where $x$ and $y$ are the training data and ground-truth labels, respectively. During this training phase, $\varepsilon_{\theta}$ is not frozen; therefore, the latent features $f$ are further adjusted to the benchmark QaTa-COV19 dataset. Overall, during the inference, ReCovNet does not require prior lung segmentation to provide reliable COVID-19 detection. Finally, we propose two versions of the proposed model: ReCovNet-v1 is formed by DenseNet-121 encoder due to its good performance in the COVID-19 detection task, and ReCovNet-v2 is formed by ResNet-50 encoder. 

\section{Experimental Evaluation}\label{sec:experiments}
In this section, the experimental setup is presented. Then, the experimental results are given on the benchmark QaTa-COV19 dataset. 

\subsection{Experimental Setup}
The performance metrics are calculated on the test (unseen) set of the QaTa-COV19 dataset. We consider COVID-19 CXRs as positive-class, whereas control group samples as negative-class. Accordingly, we form the confusion matrix (CM) elements as follows: true positive is the number of correctly classified COVID-19 samples, false positive is the number of misclassified control group samples as the positive class member, true negative is the number of correctly detected control group samples, and false negative is the number of misclassified COVID-19 samples as the negative class members. The performance metrics are defined as follows: \textit{sensitivity} is the rate of correctly detected COVID-19 samples in all positive samples, \textit{specificity} is the ratio of correctly classified control group samples in all negative samples, \textit{precision} is the rate of correctly classified positive samples among all the members detected as positive class members, \textit{accuracy} is the rate of correctly detected samples among all the data. Moreover, we define the \textit{F-score} as follows:
\begin{equation}
    F(\beta) = (1+\beta^2)\frac{(\textit{precision}\times\textit{sensitivity})}{\beta^2\times\textit{precision}+\textit{sensitivity}}
\end{equation}
where the harmonic average between \textit{precision} and \textit{sensitivity} is defined as  \textit{F$1$-score} as $\beta=1$. On the other hand, to minimize the effect of false negatives over false positives, \textit{F$2$-Score} is defined as $\beta=2$. The major performance metric in COVID-19 detection is \textit{sensitivity}  since any misdetection of the disease threatens global health. Hence minimizing the false alarm ($1-\textit{specificity}$) is our target.

\begin{table*}[t!]
\centering
\caption{COVID-19 detection performance results (\%) computed over the test (unseen data) set of QaTa-COV19 dataset using four state-of-the-art and the proposed ReCovNet deep models.}
\begin{tabular}{|c|cccccc|}
\hline
Model &  Sensitivity& Specificity & Precision & F1-Score & F2-Score & Accuracy \\ \hline \hline
 
\rowcolor[gray]{0.95}ResNet-50 & $96.571$ & $99.953$ & $98.734$ & $97.641$ & $96.996$ & $99.828$ \\

Inception-v3 & $94.762$  & $99.821$ & $95.307$  & $95.033$ & $94.870$ & $99.634$ \\

\rowcolor[gray]{0.95}Inception-ResNet-v2 & $94.286$ & $99.803$ & $94.828$ & $94.556$ & $94.394$ & $99.599$ \\

DenseNet-121 & $97.429$ & $\textbf{99.974}$ & $\textbf{99.320}$ & $\textbf{98.365}$ & $\textbf{97.801}$ & $\textbf{99.880}$ \\

\rowcolor[gray]{0.95}ReCovNet-v1 & $97.810$ & $99.901$ & $97.438$ & $97.624$ & $97.735$ & $99.824$ \\ 

ReCovNet-v2 & $\textbf{98.571}$ & $99.770$ & $94.262$ & $96.369$ & $97.678$ & $99.726$ \\ \hline

\end{tabular}
\label{tab:detection-results}
\end{table*} 

The networks are implemented using Tensorflow library on NVidia ® GeForce RTX 2080 Ti GPU card. The optimizer choice is Adam with its default momentum parameters. ReCovNet models are trained with $15$-epochs, the learning rate of $\alpha=10^{-5}$, and a batch size of $64$. The segmentation networks are trained with $15$-epochs, the learning rate of $\alpha=10^{-4}$, and a batch size of $32$. We have utilized Montgomery County X-ray Set \cite{seg1} and Japanese Society of Radiological Technology (JSRT) \cite{seg2} datasets to train the segmentation models. All the images are frontal-view CXRs and have their corresponding ground-truths except for the JSRT. Thus, the segmentation masks provided by \cite{seg3} are used as ground-truths for JSRT \cite{seg2}. Overall, the number of CXRs is $385$ in the lung segmentation dataset. For the performance evaluation, we split this data with a ratio of 80\% training to 20\% test sets. Then, training samples are augmented up to $1000$ samples. 

\subsection{Experimental Results}
In this section, the performance of segmentation networks is first investigated. Over the test set of segmentation dataset, the segmentation model with DenseNet-121 encoder has achieved $96.12\%$ sensitivity and $98.59\%$ specificity, and with ResNet-50 encoder $97.12\%$ sensitivity and $98.22\%$ specificity for the lung segmentation task.

The COVID-19 detection performance results of the state-of-the-art and ReCovNet models are presented in Table \ref{tab:detection-results}. For each model, we have observed that their performance on COVID-19 detection is successful with $>94\%$ sensitivity. The best model from the state-of-the-art deep models is DenseNet-121 with $97.43\%$ sensitivity and $99.97\%$ specificity. The performance of ReCovNet-v1 is very close to DenseNet-121. However, the best sensitivity in COVID-19 detection is achieved by the ReCovNet-v2 by $98.57\%$, which is an outstanding performance for the diagnosis on the largest COVID-19 dataset. Moreover, ReCovNet-v2 also holds a high specificity level of $99.77\%$. Table \ref{tab:CMs} shows the confusion matrices of the best performing models, which are DenseNet-121 from state-of-the-art deep models and ReCovNet-v2 from the proposed networks. The best detection (sensitivity) rate is achieved by ReCovNet-v2, which misses only $15$ COVID-19 samples among $1050$ images. 

\begin{table}[b!]
\centering
\caption{Confusion matrices of the best performing DenseNet-121 and proposed ReCovNet-v2 model for COVID-19 detection.}
\begin{subtable}{.48\textwidth}
\centering
\caption{Confusion Matrix DenseNet-121}
\begin{tabular}{|c|c|c|c|}
\hline
\multicolumn{2}{|c|}{\multirow{2}{*}{\textbf{DenseNet-121}}} & \multicolumn{2}{c|}{Predicted} \\ \cline{3-4} 
\multicolumn{2}{|c|}{} & \multicolumn{1}{c|}{Control Group} & \multicolumn{1}{c|}{COVID-19} \\ \hline
\multirow{2}{*}{\begin{tabular}[c]{@{}c@{}}Ground\\ Truth\end{tabular}} & Control Group & $27390$ & $7$ \\ \cline{2-4} 
 & COVID-19 & $27$ & $1023$ \\ \hline
\end{tabular}
\label{CMa}
\end{subtable}

\bigskip
\noindent
\begin{subtable}{.48\textwidth}
\centering
\caption{Confusion Matrix ReCovNet-v2}
\begin{tabular}{|c|c|c|c|}
\hline
\multicolumn{2}{|c|}{\multirow{2}{*}{\textbf{ReCovNet-v2}}} & \multicolumn{2}{c|}{Predicted} \\ \cline{3-4} 
\multicolumn{2}{|c|}{} & \multicolumn{1}{c|}{Control Group} & \multicolumn{1}{c|}{COVID-19} \\ \hline
\multirow{2}{*}{\begin{tabular}[c]{@{}c@{}}Ground\\ Truth\end{tabular}} & Control Group & $27334$ & $63$ \\ \cline{2-4} 
 & COVID-19 & $15$ & $1035$ \\ \hline
\end{tabular}
\label{CMb}
\end{subtable}
\label{tab:CMs}
\end{table}

\begin{figure}[t!]
    \centering
    \includegraphics[width=.48\textwidth]{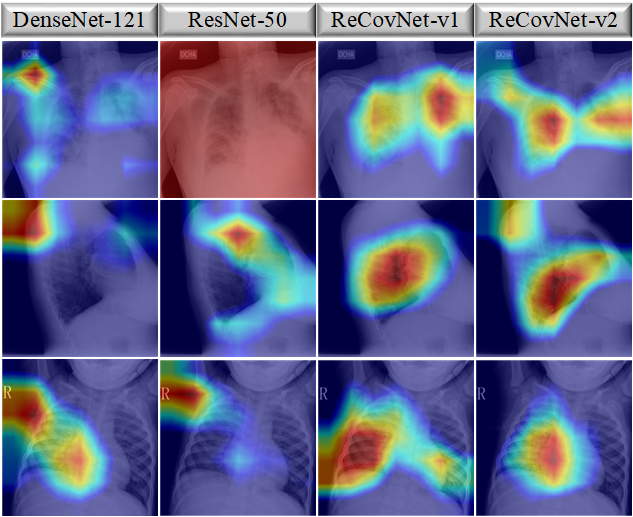}
    \caption{The activation maps are extracted by Grad-CAM \cite{selvaraju2017grad} approach for the models. The top two rows are COVID-19 samples, whereas the bottom row is a CXR from the control group images.}
    \label{fig:activations}
\end{figure}

The results on the largest COVID-19 dataset, which includes many CXR images from different thoracic diseases, shows that deep models can achieve elegant COVID-19 detection performance. However, the activation maps extracted by Grad-CAM \cite{selvaraju2017grad} approach reveal the contribution of the irrelevant regions and this is a major issue of these models in COVID-19 diagnosis. To exemplify this issue, we have compared the proposed ReCovNet-v1 and ReCovNet-v2 models with deep models as shown in Fig. \ref{fig:activations}. The activation maps show that DenseNet-121 and ResNet-50 models obviously get the information from irrelevant regions on the CXRs while the proposed models focus on the relevant regions.  

\section{Conclusions}\label{sec:conclusion}
The diagnosis of COVID-19 is a crucial task to prevent the further spread of the disease. This study investigates the limitations of the state-of-the-art deep models that are trained for COVID-19 detection directly from CXRs. To address these problems, we propose an end-to-end reliable COVID-19 detection network with pre-trained convolutional layers. We have compiled and publicly shared the largest COVID-19 dataset: QaTa-COV19, which includes $4603$ COVID-19 samples, and $120,013$ CXRs from $14$ different thoracic diseases and normal samples. The experimental results over this benchmark dataset have shown that the proposed approach has achieved the highest sensitivity level compared to competing methods. We also demonstrated how the proposed models  properly focus their analysis in the relevant region of the CXR instead of irrelevant activation observed in the competing models. In our future work, more CXR images will be used to train the lung segmentation models to further increase the reliability of our approach in COVID-19 detection.

\newpage
\bibliographystyle{IEEEbib}
\bibliography{refs}

\end{document}